\documentclass[a4paper,11pt]{article}

\usepackage{jheppub} 

\usepackage[T1]{fontenc} 

\usepackage{graphicx}
\usepackage{amsfonts}
\usepackage{braket}
\usepackage{bbm}
\usepackage{slashed}
\usepackage{fontenc}
\usepackage{microtype}
\usepackage{textcomp}
\usepackage{epstopdf}
\usepackage{pstricks}
\usepackage{color}

\newtheorem{mydef}{Definition}

\newtheorem{mytheo}{Theorem}
\newtheorem{mycorol}{Corollary}
\title{\boldmath Two-loop Integral Reduction from Elliptic and
  Hyperelliptic Curves}

\author[a]{Alessandro Georgoudis}
\author[a]{Yang Zhang}

\affiliation[a]{ETH Z\"urich, Institute for Theoretical Physics, Wolfgang-Pauli-Str. 27,
8093 Z\"urich,
Switzerland}

\emailAdd{georgoua@student.ethz.ch}
\emailAdd{yang.zhang@phys.ethz.ch}

\abstract{We show that for a class of two-loop diagrams, the on-shell
  part of the integration-by-parts (IBP) relations correspond to exact meromorphic one-forms on
  algebraic curves.  Since it is easy to find such exact
  meromorphic one-forms from algebraic geometry,
this idea provides a new highly efficient algorithm for integral reduction. We
demonstrate the power of this method via several complicated two-loop
diagrams with internal massive legs. No explicit elliptic or hyperelliptic integral
computation is needed for our method.}

\begin{document} 
\maketitle
\flushbottom
\section{Introduction}
With the beginning of Run II of the Large Hadron Collider (LHC), we
need high precision scattering amplitudes in Quantum Chromodynamics
and the Standard Model, to reduce the theoretical uncertainty. The precise scattering amplitude computation
suffers from problems of the large number of loop Feynman diagrams and difficult integrations for each loop diagram. This paper aims at
developing a new method of reducing loop integrals to the minimal set
of integrals, i.e., master integrals (MIs).

Traditionally, integral reduction can be achieved by
applying integration-by-parts (IBP) identities
\cite{Chetyrkin:1981qh} and considering the other
symmetries of loop diagrams. However, given a two-loop or higher-loop
integral, it is difficult to find a particular IBP identity
which reduce it to MIs without introducing unwanted terms. There are
several implements of IBPs generating codes \rm{AIR} \cite{
  Anastasiou:2004vj}, \rm{FIRE} \cite{Smirnov:2013dia,
  Smirnov:2008iw,Smirnov:2006tz} and Reduze \cite{vonManteuffel:2012yz, Studerus:2009ye}, based on Laporta algorithm
\cite{Laporta:2001dd}, by the computation of Gaussian elimination or
Gr\"obner basis. For multi-loop diagrams with high multiplicities or many mass scales, it may take a lot of time and computer
RAM to finish the integral reduction. There are also several new approaches
for integral reduction, based on the study of the Lie algebra structure of
IBPs \cite{Lee:2008tj}, Syzygy computation
\cite{Gluza:2010ws, Schabinger:2011dz}, reductions over finite fields
\cite{vonManteuffel:2014ixa}, and differential geometry
\cite{Zhang:2014xwa}. Besides, the number of master integrals can be
determined by the critical points of polynomials \cite{Lee:2013hzt}.

We present a new method of integral reduction, for a class of
multi-loop diagrams, based on unitarity  \cite{Bern:1994zx,Bern:1994cg,Britto:2004nc,Britto:2004ap,Britto:2004nj,Britto:2005fq, Britto:2005ha,Anastasiou:2006jv,Anastasiou:2006gt,Cachazo:2004dr, Kosower:2011ty, Johansson:2012zv, Sogaard:2013yga,
  Johansson:2013sda,CaronHuot:2012ab, Sogaard:2013fpa, Sogaard:2014jla,
  Johansson:2015ava} and the analysis of {\it algebraic
curves} \cite{griffiths2011principles,miranda1995algebraic,farkas1992riemann}. We show that for a $D$-dimensional $L$-loop diagram, if 
the unitarity cut solution $V$ is an irreducible algebraic curve,
then the on-shell IBPs of this diagram correspond to {\it exact
  meromorphic 1-forms} on $V$. For an algebraic curve, it is
very easy to find the exact meromorphic 1-froms, based on algebraic
geometry. Hence for this class of diagrams, we can derive the on-shell
part of IBPs very efficiently from our method. 

Schematically, an IBP relation,
\begin{equation}
\int \frac{d^D l_1}{(2 \pi)^{D}} \ldots \frac{d^D l_L}{(2\pi)^{D}}
\frac{\partial }{\partial l_i^\mu}\frac{v_i^\mu}{D_1^{\alpha_1} \ldots D_k^{\alpha_k}}  = 0
\end{equation}
on the unitarity cut $V: D_1 =\ldots =D_k=0$ becomes contour integral
relations \cite{Kosower:2011ty, Johansson:2012zv, Sogaard:2013yga,
  Johansson:2013sda,CaronHuot:2012ab, Sogaard:2013fpa,Sogaard:2014jla,
  Johansson:2015ava},
\begin{equation}
  \oint \omega =0,
\label{exact}
\end{equation}
where the integrals are along combinations of non-trivial cycles of 
$V$, contours surrounding singular points of $V$ and
poles of $\omega$. If $V$ is an algebraic curve, then the contours are
one-dimensional. Furthermore, if $V$ is irreducible, then $V$ has a
complex structure and $\omega$ is a meromorphic 1-form
\cite{griffiths2011principles}. In this case, (\ref{exact}) holds for
all contours, so it implies the $\omega$ is an exact form,
\begin{equation}
  \label{eq:2}
  \omega=d F,
\end{equation}
where $F$ is a meromorphic function on $V$. So the on-shell part of IBP relations for
this diagram correspond to exact meromorphic 1-forms. Mathematically,
it is very easy to find exact meromorphic 1-forms on an algebraic
curve,
so we can quickly get the on-shell IBPs for this diagram.

In this paper, we show several two-loop examples with internal masses for our new
method. Multi-loop integrals with internal masses appear frequently in
QCD/SM scattering amplitudes, and are bottlenecks for integral
reduction or evaluation. We use these complicated cases to show the
power of our method:  
\begin{enumerate}
\item $D=4$ planar double box with internal massive legs. The unitarity cut for this diagram is
  an elliptic curve. The maximal unitarity structure of the symmetric double box, with internal massive
  legs, was studied in \cite{Sogaard:2014jla}. Here we derive the integral reduction for the general
  cases, based on the analysis of differential forms on an elliptic
  curve. We also reduce integrals with doubled propagators based on
  algebraic curves,
  which were not considered in  \cite{Sogaard:2014jla}.
\item $D=2$ sunset diagram. The unitarity cut for this diagram
  is again an elliptic curve. We derive the analytic integral reduction based on the analysis of elliptic curves.
\item $D=4$ non-planar crossed box with internal massive legs.  The unitarity cut for this diagram
  is a genus-$3$ hyperelliptic curve. The unitarity cut is more
  complicated than the planar two-loop counterparts, however, our method also works for this
  case. We get the analytic integral
  reduction from the analysis of the hyperelliptic curve.
\end{enumerate}
In these examples, we get all the on-shell IBPs analytically. The
algorithm is realized by a {\rm Mathematica} code containing algebraic
geometry tools. For each diagram, the analytic integral reduction is
extremely fast, which has the time order of minutes.

We have the following remarks,
\begin{itemize}
\item Although the mathematic objects are elliptic or hyperelliptic, we do
not need the explicit form of elliptic/hyperelliptic functions, or elliptic/hyperelliptic integrals. Only the differential relations for elliptic and
hyperelliptic functions are needed. These relations involve
rational coefficients only and are easy to find.
\item The method presented in this paper is different from the maximal
  unitarity method. For the maximal unitarity method, we need to
  perform contour integrals to extract the master integral
  coefficients. Our method use the integrand reduction  \cite{Ossola:2006us,
    Ossola:2007ax,Mastrolia:2011pr,Badger:2012dv} via Gr\"obner
  basis \cite{Zhang:2012ce,
    Mastrolia:2012an,Feng:2012bm,Badger:2012dv,Mastrolia:2012wf, Mastrolia:2013kca,Badger:2013gxa} first, to reduce the
  loop amplitude to an integrand basis. Then we use the knowledge of
  algebraic curves, to reduce the integrand basis further to master
  integrals. In this way, we avoid the explicit elliptic or
  hyperelliptic integral computations.
\end{itemize}

This paper is organized as follows: In section 2, we present our
method based on algebraic curves. In section 3 and 4, the
double box diagram (elliptic) and sunset diagram (elliptic) with
internal masses
will be explicitly presented. In section 5, we consider the integral
reduction for the
massive nonplanar box diagram (hyperelliptic). The rudiments of the
knowledge of algebraic curves are included in the appendix.

\section{Integral Reduction via the Analysis of Algebraic Curves}  
Generically, for a quantum field theory, the $L$-loop amplitude can be
written as \cite{Bern:1994zx,Bern:1994cg},
\begin{equation}
A_n^\text{$L$-loop}=\sum_{k} c_k I_k + \text{rational terms}\,, 
\label{MI}
\end{equation}
The set $\{I_k\}$ is called the master integral (MI) basis whose
elements are independent loop integrals. In practice, for amplitudes with multiple loops, high multiplicities or several mass
scales, it is quite difficult to determine the set of master integral
or reduce a generic integral, 
\begin{equation}
\int \frac{d^D l_1}{(2 \pi)^{D/2}} \ldots \frac{d^D l_L}{(2 \pi)^{D/2}}
\frac{N(l_1,\ldots l_L)}{D_1^{\alpha_1} \ldots D_k^{\alpha_k}} ,
\end{equation}
to the linear combination of master integrals.

Traditionally, the integral reduction is done by using IBP identities \cite{Chetyrkin:1981qh},
\begin{equation}
\int \frac{d^D l_1}{(2\pi)^{D}} \ldots \frac{d^D l_L}{(2\pi)^{D}}
\frac{\partial }{\partial l_i^\mu}\frac{v_i^\mu}{D_1^{\alpha_1} \ldots D_k^{\alpha_k}}  = 0,
\end{equation}
if there is no boundary term. In general, it is difficult to find the
IBP relations for a multi-loop integral reduction. 

We present a new way of integral reduction, based on maximal unitarity
method and algebraic curves. Given a Feynman integral with $k$
propagators, maximal unitarity method split (\ref{MI}) as
\cite{ Britto:2004nc, Britto:2005ha, Britto:2004ap, Britto:2004nj, Britto:2005fq,
  Anastasiou:2006jv,Anastasiou:2006gt, Cachazo:2004dr, Kosower:2011ty, Johansson:2012zv,
  Johansson:2013sda,CaronHuot:2012ab,Sogaard:2013fpa, Sogaard:2013yga, Sogaard:2014jla, Johansson:2015ava} ,
\begin{gather}
\text{Int}=\sum_i c_i I_i + \big(\text{integrals with
       fewer-than-k propagators}\big)+\text{rational terms}
\label{maximal_unitarity}
\end{gather}
where the first sum is over the master integral with exact $k$ propagators. 

The condition the all internal legs are on-shell, is called the
maximal unitarity cut,
\begin{equation}
  \label{eq:1}
  V: \quad D_1=\ldots =D_k=0,
\end{equation}
and the solution set for this equation system is an {\it algebraic
  variety} $V$.  $V$ can be a set of discrete points, algebraic curves or
surfaces.  (See \cite{Huang:2013kh,Hauenstein:2014mda} for the
detailed mathematical study of multi-loop unitarity cut solutions.) Maximal unitarity replaces the original integral with
contour integrals \cite{Kosower:2011ty, Johansson:2012zv, Sogaard:2013yga,
  Johansson:2013sda,CaronHuot:2012ab, Sogaard:2013fpa,Sogaard:2014jla,
  Johansson:2015ava}, schematically, 
\begin{eqnarray}
  \label{eq:3}
  \int \frac{d^D l_1}{(2\pi)^{D}} \ldots \frac{d^D l_L}{(2\pi)^{D}}
\frac{N(l_1,\ldots l_L)}{D_1^{\alpha_1} \ldots D_k^{\alpha_k}}
&\rightarrow& \oint \frac{d^D l_1}{(2\pi)^{D}} \ldots \frac{d^D l_L}{(2\pi)^{D}}
\frac{N(l_1,\ldots l_L)}{D_1^{\alpha_1} \ldots
              D_k^{\alpha_k}}\nonumber \\
&=& \sum_j w_j \oint_{\mathcal C_j} \omega
\end{eqnarray}
where $\omega$ is a differential form on $V$, and contours $c_j$'s are
around the poles of $\omega$ and also the {\it nontrivial cycles} of
$V$ \cite{CaronHuot:2012ab, Sogaard:2014jla}. $w_j$ are weights of
these contours. In particular, to extract the
coefficients $c_i$ in (\ref{MI}), we can find a special set of weights
$w_j^{\{i\}}$ \cite{Kosower:2011ty, Johansson:2012zv,
  Johansson:2013sda,CaronHuot:2012ab,Sogaard:2013yga,Sogaard:2013fpa, Sogaard:2014jla} such that,
\begin{equation}
  c_i =  \sum_j w_j^{\{i\}} \oint_{\mathcal C_j} \omega
\label{linear_fitting_c}
\end{equation}
For example, the $4D$ two-loop massless double box diagram contains $7$
propagators. The maximal cut is a reducible variety with $6$
components  \cite{Kosower:2011ty}, each of which is a Riemann sphere. The contours are around
the intersecting points of these Riemann spheres, which are singular
points of this variety, and also around the poles of $\omega$. For the
$4D$ two-loop double box with massive internal legs, the
maximal cut gives one irreducible variety, which is an elliptic
curve \cite{Sogaard:2014jla}. There is no singular point on this variety, so the contours are around the poles
of $\omega$ and also the {\it two fundamental cycles} on the elliptic
curve.

Our observation is that if a differential form $\omega$ on $V$ is
integrated to zero, around all singular points on $V$, poles of
$\omega$ and non-trivial cycles of $V$. 
\begin{equation}
  \label{form_zero_integral}
  \oint_{\mathcal C_j} \omega=0, \quad \forall j 
\end{equation}
then from (\ref{linear_fitting_c}) and (\ref{maximal_unitarity}), the integral corresponding to
$\omega$ can be reduced to integrals with fewer propagators. Since
from the knowledge of algebraic geometry, it is easy to find such
$\omega$'s satisfying (\ref{form_zero_integral}), we propose a new
multi-loop integral reduction method from this viewpoint. 

In this paper, we focus on the cases for which the
number of propagators equals $DL-1$ and the maximal unitarity cut
gives one irreducible variety. In such a case, the cut solution $V$
is a smooth {\it algebraic curve} with well defined complex
structure. The condition (\ref{form_zero_integral}) implies that
$\omega$ is an exact meromorphic form on $V$, since the integral
\begin{equation}
  \label{eq:6}
  F(P)=\int_O^P \omega, \quad \forall P\in V
\end{equation}
is independent of the path and $dF=\omega$. Then from the study of
meromorphic functions on $V$, which is a well-known branch of
algebraic geometry, we can list generators for $F$ and then derive all
forms which satisfy  (\ref{form_zero_integral}).

Explicitly, for this class of diagram, we found that the scalar integral on the cut becomes a holomorphic form on $V$. 
\begin{equation}
\int \frac{d^D l_1}{(2\pi)^{D}} \ldots \frac{d^D l_L}{(2\pi)^{D}}
\frac{1}{D_1 \ldots D_k} \bigg|_\text{cut}= \oint \Omega
\label{scalar_cut}
\end{equation}
where the $1$-form $\Omega$ is globally holomorphic (without poles) on
$V$. On the cut, the components of $l_i$'s become meromorphic
functions. We can show that these functions generate all meromorphic
functions on $V$. Let $F(l_1,\ldots l_L)$ be a polynomial in the
components of loop momenta, then take the derivative of $F$,
\begin{equation}
  dF = f \Omega.
\label{exteriord}
\end{equation}
The resulting $f \Omega$ is an exact meromorphic $1$-form. From the analysis
above, we get that,
\begin{equation}
  \label{eq:9}
  \int \frac{d^D l_1}{(2\pi)^{D}} \ldots \frac{d^D l_L}{(2\pi)^{D}}
\frac{f}{D_1 \ldots D_k} =0+  (\text{integrals with
       fewer propagators}),
\end{equation}
so we obtain an integral reduction relation. For the explicit examples
in this paper, we can show that this method provides all the on-shell
part of integral reduction relations. 

In practice, our algorithm can be presented as,
\begin{enumerate}
\item Use integrand reduction method via Gr\"obner basis
  \cite{Zhang:2012ce,Mastrolia:2012an} to rewrite the loop scattering
  amplitude as the form of {\it integrand basis}. The coefficients of
  integrand basis can be determined by fusing tree amplitudes or
  polynomial division of the Feynman integrand.
\item Calculate the maximal cut of the scalar integral to determine
  the form of holomorphic form $\Omega$ as (\ref{scalar_cut}).
\item Calculate the exterior derivatives of all polynomial $F$'s. In practice, it is sufficient to consider the linear
  $F$'s and then use the chain rule. Then we get all the on-shell IBP
  relations as (\ref{exteriord}).
\item Use the obtained integral reduction relations to reduce the
  integrand basis to master integrals.
\end{enumerate}
If the integrand contains doubled-propagator integrals, the algorithm
will be slightly different. We need to solve a polynomial Diophantine equation
first, and the procedure will be shown in the next section.

Note that our algorithm is different from the traditional maximal
unitarity. Usually, maximal unitarity method needs the explicit
contour integration to extract the master integral coefficients. However, our algorithm does not require the
explicit contour integration, and the explicit form of elliptic/hyperelliptic
integral is not needed. The residue computations to find the
holomorphic form $\Omega$ and the derivative computations
(\ref{exteriord}) are much simpler than the contour integrals.

\section{Elliptic Example: Double Box with Internal Masses}
\label{sec:double box}
\begin{figure}[ht]
\begin{center}
\includegraphics{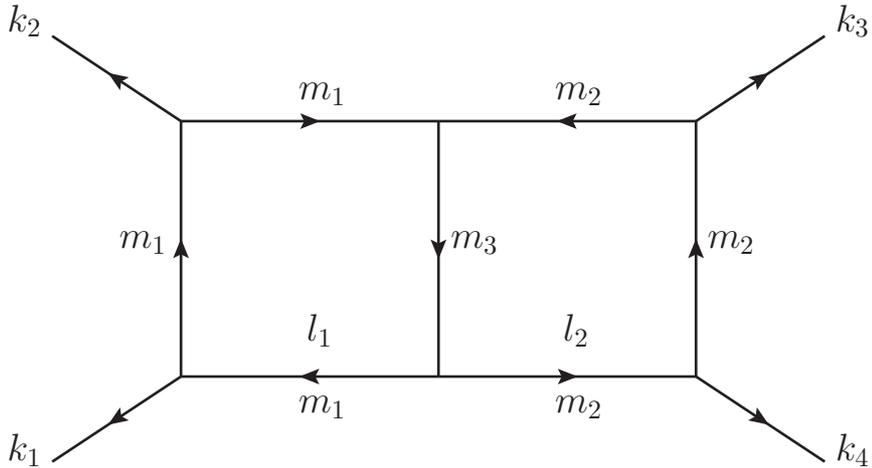}
\end{center}
\caption{Planar double box diagram with $3$ internal mass scales}\label{planarb}
\end{figure}

The method explained in the previous section can be used for integral
reduction for various topologies, for instance, the double-box (Fig. \ref{planarb}) with three different
masses for the internal propagators. The maximal unitarity of the
massless double box was discussed in \cite{Kosower:2011ty}. Then,
maximal unitarity
structures for double box with $1\sim 4$ massive external legs were studied in 
\cite{Johansson:2012zv,Johansson:2013sda}. In these cases, the
unitarity cuts provide reducible curves. 

On the other hand, the unitarity cut of double box with six massive
external legs \cite{CaronHuot:2012ab} or all
massive internal legs provides irreducible curves. The integral reduction for
symmetric double box diagram with internal masses, was discussed in
\cite{Sogaard:2014jla}, via maximal unitarity and the analysis of elliptic
functions. Here we show the integral reduction for more
generic double box diagram with $3$ internal mass scales, based on our new
method, without using elliptic functions/integrals explicitly. 

\subsection{Maximal unitarity}
The denominators for double box diagrams are
 \begin{equation}
\label{denominators}
\begin{split}
D_1 &=l_1^2-m_1^2 \,,
\qquad
D_2 =(l_1-k_1)^2-m_1^2  \,,
\qquad
D_3 =(l_1-k_1-k_2)^2-m_1^2  \,,
\\
D_4 &=l_2^2-m_2^2 \,,
\qquad
D_5 =(l_2-k_4)^2-m_2^2  \,,
\qquad
D_6 =(l_2-k_3-k_4)^2-m_2^2  \,,
\\F
D_7&=(l_1+l_2)^2-m_3 \,.
\end{split}
\end{equation}

We parametrize the loop momenta as,
\begin{equation}
\begin{aligned}
l_1^{\mu} & =\alpha_1 k_1^{\mu}+ \alpha_2 k_2^{\mu}+ \alpha_3 \frac{s}{2} \frac{\bra{1} \gamma^{\mu} | 2]}{\braket{1 4}[4 2]}+ \alpha_4 \frac{s}{2} \frac{\bra{2} \gamma^{\mu} | 1]}{\braket{2 4}[4 1]}, \\ 
l_2^{\mu} & =\beta_1 k_3^{\mu}+ \beta_2 k_4^{\mu}+ \beta_3 \frac{s}{2} \frac{\bra{3} \gamma^{\mu} | 4]}{\braket{3 1}[1 4]}+ \beta_4 \frac{s}{2} \frac{\bra{4} \gamma^{\mu} | 3]}{\braket{4 1}[1 3]},
\end{aligned}
\label{dbox_Para}
\end{equation}
and the Jacobian for this parameterization is,
\begin{equation}
\label{Para_Jacobian}
J_1= \underset{\mu,i}{\mathrm{det}}\frac{\partial l_1^{\mu}}{\partial \alpha_i}=\frac{\mathrm i s^4}{4 t (s + t)}\,, \qquad J_2= \underset{\mu,i}{\mathrm{det}}\frac{\partial l_2^{\mu}}{\partial \beta_i}=\frac{\mathrm i s^4}{4 t (s + t)}\,.
\end{equation}

The solutions for the maximal unitarity cut,
\begin{equation}
  \label{eq:1}
  D_1=D_2=\ldots = D_7=0.
\end{equation}
defines an elliptic curve. To see this, we first solve for the
variables $\alpha_1$, $\alpha_2$, $\alpha_3$, $\beta_1$, $\beta_2$ and
$\beta_3$ in terms of $\alpha_4$ and $\beta_4$,
 \begin{equation}
\label{soludouble}
\begin{split}
\alpha_1 &= 1 \,,
\qquad
\alpha_2 = 0 \,, \qquad
\alpha_3 = \frac{m_1^2 t (s + t)}{
 \alpha_4 s^3},
\\
\beta_1 &=0\,,
\qquad
 \beta_2 = 1\,, \qquad
 \beta_3 =\frac{m_2^2 t (s + t)}{
 \beta_4 s^3}\,, \\
 \end{split}
\end{equation}
Then the remaining one equation relates $\alpha_4$ and $\beta_4$,
\begin{equation}
  \label{elliptic_curve_dbox}
  K(\alpha_4,\beta_4)=A(\alpha_4) \beta_4^2+ B(\alpha_4)
  \beta_4+C(\alpha_4) =0,
\end{equation}
Here $A(\alpha_4)$, $B(\alpha_4)$ and $C(\alpha_4)$ are quadratic polynomials of
$\alpha_4$, whose coefficients depend on kinematic
variables. Formally, $\beta_4$ depends on $\alpha_4$ as,
\begin{equation}
  \label{db_formal_1}
  \beta_4 = \frac{-B(\alpha_4) \pm
    \sqrt{\Delta(\alpha_4)}}{2A(\alpha_4)}\, , \quad \Delta=B^2-4AC,
\end{equation}
where $\Delta$ is a quartic polynomial in $\alpha_4$ with {\it four
  distinct roots}, for generic kinematics with internal masses. Hence
the maximal unitarity cut defines an elliptic curve, i.e., algebraic curve with genus one,
\begin{equation}
  \label{eq:5}
  \mathcal C:       \eta^2 =\Delta(\alpha_4),  
\end{equation}
(See the appendix for the basis introduction to elliptic curves.) The choice of keeping $\alpha_4$ and $\beta_4$ and eliminating other
variables is purely arbitrary. 

On the cut, by a short calculation, the scalar double box integral,
\begin{equation}
I=\int \frac{d^4 l_1}{(2 \pi)^4} \frac{d^4 l_2}{(2 \pi)^4} \frac{1}{D_1\ldots D_7}\,,
\end{equation}
has the following structure:
\begin{equation}
I|_{7-cut} =\frac{s^2 t}{16} \oint \frac{d\alpha_4}{\sqrt{\Delta}}
\end{equation}
where the overall factor is not important for the following discussion. 

As \cite{Sogaard:2014jla}, it is remarkable that
$\frac{d\alpha_4}{\sqrt{\Delta}}$ is the only {\it holomorphic one-form}
associated with the elliptic curve. On the cut, the loop-momentum
components $\alpha_i$, $\beta_i$ become elliptic functions. So we may
study the explicit form of these functions, calculate the elliptic
integrals and perform the integral reduction, in a procedure of  \cite{Sogaard:2014jla}.
However, in this paper, we propose a different procedure: (1) reduce
the integrand based on Gr\"obner basis \cite{Zhang:2012ce, Mastrolia:2012an} (2) reduce
the  integrand basis to master integrals by the study of
differential forms on the elliptic curve. The advantage of this
approach is that the whole computation is very simple: no explicit
elliptic parameterization or elliptic integral is needed. The new
process can also be easily automated on computer algebra systems.


\subsection{Integral reduction}
\label{ibpmibdb}
We now focus on the double box integral with numerator $N$, 
\begin{equation}
I[N]=\int \frac{d^4 l_1}{(2 \pi)^4} \frac{d^4 l_2}{(2 \pi)^4} \frac{N}{D_1\ldots D_7}\,,
\end{equation}

Integrand reduction method via Gr\"obner basis
method \cite{Zhang:2012ce,Mastrolia:2012an} determines that the integrand basis contains
$32$ terms. In terms of (\ref{dbox_Para}), the basis can be presented
as, 
\begin{gather}
\mathcal B=\{ \alpha_3^4 \beta_3,\alpha_4 \beta_4^4,\alpha_4
   \beta_3^4,\alpha_4^4 \beta_3,\beta_4^4,\beta_3^4,\alpha_3^3 \beta_3,\alpha_3^4,\alpha_4 \beta_4^3,  
    \alpha_4
   \beta_3^3,\alpha_4^3
   \beta_3,\alpha_4^4,\beta_4^3,\beta_3^3,\alpha_3^2
   \beta_3,\alpha_3^3,\nonumber \\\alpha_4 \beta_4^2,\alpha_4
   \beta_3^2,\alpha_4^2 \beta_3,\alpha_4^3,\beta_4^2,\beta_3^2,\alpha_3 \beta_3,\alpha_3^2, 
    \alpha_4 \beta_4,\alpha_4
   \beta_3,\alpha_4^2,\beta_4,\beta_3,\alpha_3,\alpha_4,1 \}
\label{dbox_integrand_basis}
\end{gather}

On the cut, the integral becomes a meromorphic one-form,
\begin{equation}
  \label{dbox_cut}
  I[ N]|_\text{cut} \propto  \oint
  \frac{d\alpha_4}{\eta} N(\alpha_3,\alpha_4,\beta_3,\beta_4)
\end{equation}
where $N$ is a polynomial in $\alpha_3$, $\alpha_4$, $\beta_3$ and
$\beta_4$, and therefore also an elliptic function. We now perform the integral
reduction, following the strategy as in the previous section. The task
is to find {\it exact meromorphic one-forms} $\omega$ on this elliptic
curves. If two integrals on the cut, differ by the contour integrals of
such an $\omega$, then
\begin{equation}
  \label{eq:4}
  I[N_1]-I[N_2]|_\text{cut} =\oint \omega= 0                                    
\end{equation}
where the second equality holds for all contours, i.e., two
fundamental cycles and small contours around the poles, because
$\omega$ is exact. Then the integral reduction between $I[N_1]$ and
$I[N_2]$ is achieved at the level of double box diagram,
\begin{equation}
  \label{eq:7}
  I[N_1]-I[N_2]=0 +(\text{integrals with $< 7$ propagators} )
\end{equation}

Note the $\alpha_4$ and $\beta_4$ generate {\it all} elliptic
  functions on this elliptic curve, as shown in the appendix, (\ref{generators_hyperelliptic}).

In practice, we find that to find such $\omega$'s, it is sufficient to
consider the exterior derivatives of polynomials in $\alpha_3$, $\alpha_4$, $\beta_3$ and
$\beta_4$,
\begin{eqnarray}
  \label{dbox_seed}
  d F(\alpha_3,\alpha_4,\beta_3,\beta_4) = \frac{\partial F}{\partial
  \alpha_3} d \alpha_3 + \frac{\partial F}{\partial
  \alpha_4} d \alpha_4+\frac{\partial F}{\partial
  \beta_3} d \beta_3+\frac{\partial F}{\partial
  \beta_4} d \beta_4 \equiv f \frac{d \alpha_4}{\eta}
\end{eqnarray}
So we need to find the one forms
$\{d\alpha_3,d\alpha_4,d\beta_3,d\beta_4\}$ and then use the chain
rule to generate integral reduction relations. We can start by
calculating $d\alpha_4$ in terms of the holomorphic one-form,
\begin{equation}
d\alpha_4=\eta \frac{d\alpha_4}{\eta} =(2A(\alpha_4) \beta_4+B(\alpha_4)) \frac{d\alpha_4}{\eta}\,,
\end{equation}
where we used  the definition $\eta=\sqrt{\Delta}$ and
(\ref{db_formal_1}) to rewrite $\eta$ in function of $\beta_4$.
The purpose of this step is to get the a polynomial form of $f$.

We can now easily find $d\alpha_3$,
\begin{equation}
d\alpha_3=d\left(\frac{\lambda_1}{\alpha_4}\right)=-\lambda_1 \frac{1}{ \alpha_4^2}d\alpha_4=
-\frac{\alpha_3^2}{\lambda_1} d\alpha_4\,,\quad \lambda_1\equiv
\frac{m_1^2 t(s+t)}{s^3}
\end{equation}
the constant $\lambda_1$ is the product of $\alpha_3 \alpha_4$ on the cut.
To generate the remaining 1-forms, we again use the form of elliptic
curve. Recall that,
\begin{equation}
K(\alpha_4,\beta_4)=A(\alpha_4)\beta_4^2+B(\alpha_4)\beta_4+C(\alpha_4)=0\,.
\end{equation}
The identity $dK=0$ reads,
\begin{equation}
d\beta_4=-\left(A^{\prime}(\alpha_4)\beta_4^2+B^{\prime}(\alpha_4)\beta_4+C^{\prime}(\alpha_4) \right)\frac{d\alpha_4}{\eta}\,.
\end{equation}
Finally we can easily calculate $d\beta_3$,
\begin{equation}
d\beta_3=d\left(\frac{\lambda_2}{\beta_4} \right)=-\lambda_2 \frac{1}{\beta_4^2}d\beta_4=-\frac{\beta_3^2}{\lambda_2}d\beta_4\,,\quad \lambda_2\equiv
\frac{m_2^2 t(s+t)}{s^3}
\end{equation}
Then use the chain rule, we get all the on-shell IBPs. 
For example, from (\ref{dbox_seed}), we analytically obtain this relation,
\begin{eqnarray}
I_\text{dbox}[\alpha_4^3]&=&\frac{1}{2 s^4(4m_2^2-s)}\bigg(3 s^3 \left(m_1^2 s-m_2^2 s-m_3^2
                             s-4 m_2^2 t+s t\right)
                             I_\text{dbox}[\alpha_4^2] \nonumber\\
&+& s (4 m_1^2 s^2 t-2 m_2^2 s^2 t-2 m_3^2
   s^2 t+m_1^4 s^2-2 m_2^2 m_1^2 s^2-2 m_3^2 m_1^2 s^2+m_2^4 s^2
    +m_3^4 s^2 \nonumber\\ & &-2 m_2^2 m_3^2 s^2+2 m_1^2 s t^2-4
   m_2^2 s t^2-8 m_2^2 m_1^2 s t-8 m_2^2 m_1^2 t^2+s^2 t^2)
                               I_\text{dbox}[\alpha_4] \nonumber\\
&+&m_1^2 t (s+t) \left(m_1^2 s-m_2^2 s-m_3^2
   s-4 m_2^2 t+s t\right)  I_\text{dbox}[1] \bigg)   +\ldots 
\end{eqnarray}
where $\ldots$ stands for integrals with fewer than $7$ propagators.
Consider all polynomials whose exterior derivative satisfy the
renormalizability conditions, we obtain $23$ integral
relations. Furthermore, consider Levi-Civita insertions which integrate to zero, 
\begin{equation}
\epsilon(l_2,k_1,k_2,k_3)  ~l_2 \cdot k_1\,, \quad \epsilon(l_2,k_1,k_2,k_3) ~ l_1 \cdot k_4\,, \quad \epsilon(l_1,l_2,k_1,k_2)\,, \quad
 \epsilon(l_1,l_2,k_1,k_3)\,.
\end{equation}
we get 4 more integral relations. So the total number of MIs is,
\begin{equation}
  \label{eq:11}
  \# \text{MI}_\text{dbox}= 32-23-4=5
\end{equation}
and explicitly the MIs can be chosen as,
\begin{equation}
\text{MI}_\text{dbox}=\left\{I_\text{dbox}[\alpha_4\beta_3],I_\text{dbox}[\alpha_4^2],I_\text{dbox}[\alpha_4],I_\text{dbox}[\beta_3] ,I_\text{dbox}[1]  \right\}\,.
\end{equation}
or in the conventional choice with $X\equiv (l_1+k_4)^2/2$ and $Y\equiv (l_2+k_1)^2/2$,
\begin{equation}
\text{MI}_\text{dbox}=\left\{I_\text{dbox}[X Y],I_\text{dbox}[X^2],I_\text{dbox}[X ],I_\text{dbox}[Y] ,I_\text{dbox}[1]  \right\}\,.
\end{equation}
and for instance, the integral reduction in this basis becomes,
\begin{gather}
I_\text{dbox} [X^3] =\frac{1}{16 s \left( 4 m_2^2-s \right)} \big(
  I_\text{dbox}[1] \left(8 m_1^6 m_2^2 s-m_1^4 \left(m_2^2 \left(s^2+4
        s t+16 t^2\right)+m_3^2 s^2\right)+ \nonumber
  \right.  \\  
\left. \left. +m_1^2 s
   \left(-m_2^4 s +2 m_2^2 \left(m_3^2 s+t (s+4 t)\right)+m_3^2
     \left(2 t (s+2 t)-m_3^2 s\right)\right)+ \nonumber
 \right. \right. 
\\ \left. \left. -s^2 t
   \left(m_2^4+m_2^2 \left(t-2 m_3^2\right)+m_3^2 \left(m_3^2+t\right)\right)\right)+2 I_\text{dbox} [X] \left(m_1^4 s
   \left(s-24 m_2^2\right)+ \nonumber \right. \right. 
\\ \left. \left. +2 m_1^2 \left(2 m_2^2 \left(s^2-2 s t+4 t^2\right)+s \left(2 m_3^2 s-t (s+2 t)\right)\right)+s
   \left(m_2^4 s-2 m_2^2 \left(m_3^2 s+2 t (t-s)\right)+ \nonumber
   \right. \right. \right. \\ 
 \left. \left.  +s \left(m_3^4+4 m_3^2 t+t^2\right)\right)\right)-12
   I_\text{dbox}[X^2] s \left(m_1^2
     \left(s-8 m_2^2\right)+m_2^2 (s-4 t)+s
     \left(m_3^2+t\right)\right)  \big)\nonumber \\
+ \ldots
\label{dbox_eample}
\end{gather}
The whole
computation takes about $120$ seconds with our \rm{Mathematica}
code. The relations are numerically verified by \rm{FIRE} \cite{Smirnov:2013dia,Smirnov:2008iw}.

\subsection{Reduction of the double-propagator integrals}
One issue not discussed in \cite{Sogaard:2014jla} is the reduction
of integral with internal mass and {\it doubled propagators}. For the
double box diagram, the doubled-propagator integral on the cut also
becomes meromorphic 1-forms, so we may carry out the maximal unitarity
analysis as that in \cite{Sogaard:2014jla}. However, in this section,
we show that, our new method is more efficient for reducing these
integrals. 

Consider the diagram in Fig. \ref{planarb} with the middle propagator
doubled,
\begin{equation}
  \label{dbox_dp}
  I_{\text{dbox},2}=\int \frac{d^4 l_1}{(2 \pi)^4} \frac{d^4 l_2}{(2 \pi)^4}
  \frac{1}{D_1 D_2 D_3 D_4 D_5 D_6 D_7^2}\,,
\end{equation}
on the cut, by the degenerate residue computation with {\it
  transformation law} or {\it Bezoutian matrix computation}
\cite{Sogaard:2014ila, Sogaard:2014oka}, we have,
\begin{equation}
  \label{eq:13}
  I_{\text{dbox},2}|_{7-cut}=-\frac{s^6t^2}{16}\oint \frac{\alpha_4 B(a_4) d\alpha_4}{\Delta^{3/2}}.
\end{equation}
Unlike the one-forms in the previous subsection, here the one-form
have the denominator $\Delta^{3/2}$. It implies that we need to find
exact 1-forms like $d(F/\Delta^{1/2})$, where $F$ is a polynomial in
the loop-momenta components. 

Note that $\Delta(a_4)$ has four distinct roots, hence $\Delta(\alpha_4)$
and $\Delta'(\alpha_4)$ have no common root. By B\'ezout's identity, 
\begin{equation}
  \label{eq:12}
  \langle\Delta(a_4), \Delta'(a_4)\rangle=\langle 1 \rangle
\end{equation}
 and the polynomial Diophantine equation 
 \begin{equation}
   \label{eq:14}
   f_1(\alpha_4) \Delta(\alpha_4)+ f_2(\alpha_4)
   \Delta'(\alpha_4)=\alpha_4 B(\alpha_4)
 \end{equation}
has solutions. Such polynomials $f_1$ and $f_2$ can be explicitly
found by Euclidean division or Gr\"obner basis method. The exterior
derivative,
\begin{equation}
  \label{eq:15}
  d\bigg(\frac{-2f_2}{\Delta^{1/2}}\bigg)=\frac{f_2
    \Delta'}{\Delta^{3/2}}d\alpha_4-2 \frac{f_2'}{\Delta^{1/2}}
  d\alpha_4
\end{equation}
determines that,
\begin{equation}
  \label{dbox_double_reduction}
    I_{\text{dbox},2}|_{7-cut}=-\frac{s^6t^2}{16}\oint \frac{(f_1+2f_2') d\alpha_4}{\Delta^{1/2}}.
\end{equation}
after integrating out the exact form. Now the term $\Delta^{3/2}$ is
removed and we can reduce this integral using the result from the
previous subsection. In practice, we find a solution such that
$f_1+2 f_2'$ is a quadratic polynomial in $\alpha_4$, so at the level
of the double box,
\begin{equation}
  \label{eq:17}
  I_{\text{dbox},2}=c_0  I_{\text{dbox}}[1]+c_1
  I_{\text{dbox}}[X]+c_2  I_{\text{dbox}}[X^2] + \ldots 
\end{equation}
The three coefficients $c_0$, $c_1$ and $c_2$ are analytically found by
our method and numerically verified by \rm{FIRE}
\cite{Smirnov:2013dia,Smirnov:2008iw}.

\section{Elliptic Example: Sunset Diagram}
The sunset diagram is a two-loop diagram which attracts a lot of research
interests \cite{Broadhurst:1993mw, Berends:1993ee, Bauberger:1994nk,
  Bauberger:1994by, Bauberger:1994hx,
  Caffo:1998du, Groote:1998wy, Groote:1998ic, Caffo:2002ch,Laporta:2004rb,Pozzorini:2005ff,
  Groote:2005ay,Kniehl:2005bc,
  Groote:2012pa,Bailey:2008ib,Caffo:2008aw,Kalmykov:2008ge, MullerStach:2011ru,Adams:2013nia,Remiddi:2013joa,
  Bloch:2013tra,Adams:2014vja,Adams:2015gva}. The sunset diagram with
$3$ different masses is the simplest loop
diagram which cannot be expresses in multiple polylogarithms. 

\begin{figure}[ht]
\begin{center}
\includegraphics{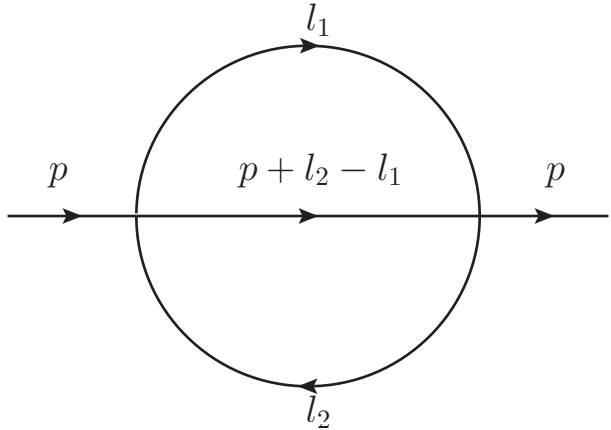}
\end{center}
\caption{Sunset diagram}\label{sundia}
\end{figure}

We use our method to study the integral reduction of the sunset
diagram (Fig. \ref{sundia}) in two dimensional space-time. In this
example, we assume that
all internal propagators are massive.

Let $p$ be the external momentum, $p^2=m^2$. We can parametrize the
loop momenta using a variant of Van-Neerven Vermaseren
 basis \cite{vanNeerven:1983vr}. Define two null vectors $e_1$ and $e_2$ such that $e_1^2=0$,
 $e_2^2=0$ and $e_1 \cdot e_2=p^2$. The Gram matrix of $\{e_1,e_2\}$ is, 
\begin{equation}
G=\begin{pmatrix}
  0 & m^2 \\
   m^2 & 0 \\
 \end{pmatrix}
\end{equation}
In this basis we expand $p$ as, 
\begin{equation}
p=e_1+\frac{e_2}{2},
\end{equation}
and define the auxiliary vector $\omega$,
\begin{equation}
\omega=e_1-\frac{e_2}{2}\,.
\end{equation}
Hence $p\cdot \omega=0$. The two loop momenta can be then generally parametrized as
\begin{equation}
\begin{split}
l_1&=\alpha_1 e_1+\alpha_2 e_2\,, \\
l_2&=\beta_1 e_1+\beta_2 e_2\,,
\end{split}
\end{equation}

On-shell equations are $D_1=D_2=D_3=0$, where $D_i$ represent the inverse propagators,
\begin{equation}\label{sundenom}
D_1 =l_1^2-m_1^2 \,,
\qquad
D_2 =(p+l_2-l_1)^2-m_2^2  \,,
\qquad
D_3 =l_2^2-m_3^2  \,.
\end{equation}
The on-shell solution can be formally expressed as,
\begin{equation}\label{solusun}
\alpha_1 = \frac{m_1^2}{2 \alpha_2 p^2} \,,
\qquad
\beta_1 =\frac{m_3^2}{2 \beta_2 p^2}\,,
\qquad
 \beta_2 = \frac{-B(\alpha_2) \pm \sqrt{\Delta(\alpha_2)}}{2A(\alpha_2)}\,.
\end{equation}
where again $\alpha_2$ and $\beta_2$ satisfy the equation of an
elliptic curve $A(\alpha_2) \beta_2^2+B(\alpha_2)
\beta_2+C(\alpha_2)=0$. The discriminant is $\Delta=B^2-4AC$. 

The sunset integral, on the triple cut, becomes contour integrals of
holomorphic 1-forms
\begin{equation}
I|_{3-cut} \propto \oint \frac{d \alpha_4}{\eta},\quad \eta\equiv \sqrt{\Delta(\alpha_2)}
\end{equation}

The integrand basis for the sunset diagram, obtained from Gr\"obner
basis method \cite{Zhang:2014xwa,Mastrolia:2012an}, contains 12 terms.
\begin{equation}\label{basissun}
\left\{ \alpha_1^2,\alpha_2^2,\alpha_1 \beta_1,\beta_1^2,\alpha_2 \beta_1,\alpha_2 \beta_2,\beta_2^2,
   \alpha_1,\alpha_2,\beta_1,\beta_2,1 \right\}
\end{equation}

We now consider the integral reduction for Fig.
\ref{sundia}, since the structure is elliptic we follow the same
strategy as the double box case. As in the double box case, we find the one forms
$\{d\alpha_1,d\alpha_2,d\beta_1,d\beta_2\}$ and then use the
chain rule to generate all the IBP relations.  We start by calculating $d\alpha_2$,
\begin{equation}
d\alpha_2=\frac{\eta}{\eta} d\alpha_2=(2A(\alpha_2) \beta_2+B(\alpha_2)) \frac{d\alpha_2}{\eta}\,,
\end{equation}
we have used (\ref{solusun}) to rewrite $\eta$ in function of $\beta_2$.
We can now easily find $d\alpha_1$,
\begin{equation}
d\alpha_1=d\left(\frac{\lambda_1}{\alpha_2}\right)=-\lambda_1 \frac{1}{ \alpha_2^2}d\alpha_2=
-\frac{\alpha_1^2}{\lambda_1} d\alpha_2,\quad
\lambda_1\equiv\frac{m_1^2}{2  m^2}
\end{equation}
Then,
\begin{equation}
d\beta_2=-\left(A^{\prime}(\alpha_2)\beta_2^2+B^{\prime}(\alpha_2)\beta_2+C^{\prime}(\alpha_2) \right)\frac{d\alpha_2}{\eta}\,.
\end{equation}
Again we can easily calculate $d\beta_1$ 
\begin{equation}
d\beta_1=d\left(\frac{\lambda_2}{\beta_2} \right)=-\lambda_2 \frac{1}{\beta_2^2}d\beta_2=-\frac{\beta_1^2}{\lambda_2}d\beta_2,\quad
\lambda_2\equiv\frac{m_3^2}{2  m^2}
\end{equation}
To generate the 1-forms, given a function $F$ which is a polynomial in
$\alpha_1$, $\alpha_2$, $\beta_1$ and $\beta_2$, we can use 
\begin{equation}
d F=\frac{\partial F}{\partial \alpha_1}d\alpha_1+\frac{\partial F}{\partial \alpha_2}d\alpha_2+\frac{\partial F}{\partial \beta_1}d\beta_1+\frac{\partial F}{\partial \beta_2}d\beta_2
\end{equation}
to generate the on-shell part of IBPs. Furthermore, note that the Levi-Civita insertions
\begin{equation}
l_1\cdot\omega\,, \quad l_2\cdot\omega\ \,, \quad l_1\cdot\omega\
l_1\cdot p, \quad l_2\cdot\omega\ l_1\cdot p \,.
\end{equation}
are integrated to zero. In total, we generate 4 IBPs and 4 independent
Levi-Civita insertions integral
relations. Hence, there are $12-4-4=4$ master integrals for the sunset
diagram. Define that $X=l_1 \cdot p$ and $Y=l_2 \cdot p$, the four
master integrals can be chosen as, 
\begin{equation}
\text{MI}_\text{sunset}=\left\{I_\text{sunset}[1], I_\text{sunset}[X],
  I_\text{sunset}[X^2], I_\text{sunset}[Y] \right\}\,.
\label{sunset_master}
\end{equation}
For instance, the reduction reads,
\begin{eqnarray}
  I_\text{sunset}[X Y]&=&\frac{1}{4} \left(m^4+m_1^2 m^2-m_2^2 m^2+m_3^2
  m^2\right)   I_\text{sunset}[1] \nonumber \\
 &+& \frac{1}{4} \left(-3
   m^2-m_1^2+m_2^2-m_3^2\right)  I_\text{sunset}[X] \nonumber \\
 &+& \frac{1}{2}  I_\text{sunset}[X^2] + \frac{1}{2}
     \left(m^2+m_1^2\right) I_\text{sunset}[Y]  +\ldots 
\end{eqnarray}
where $\ldots$ stands for integrals with fewer than $3$ propagators.

Note that generically, $D$-dimensional sunset diagrams with $3$
distinct masses have $4$
master integral. The four master integrals can be chosen as
(\ref{sunset_master}) or the scalar integral and three
doubled-propagator integrals. There is a subtlety that if $D=2$, then
the $4$ master integrals are related by Schouten identities
\cite{Remiddi:2013joa}. These identities are valid for $D<3$, and at
$D=2$ they further reduce the number of master integrals from $4$ to
$2$.   

\section{Hyperelliptic Example: Nonplanar Crossed Box with Internal Masses}
\label{sec:cross box}
\begin{figure}[ht]
\begin{center}
\includegraphics{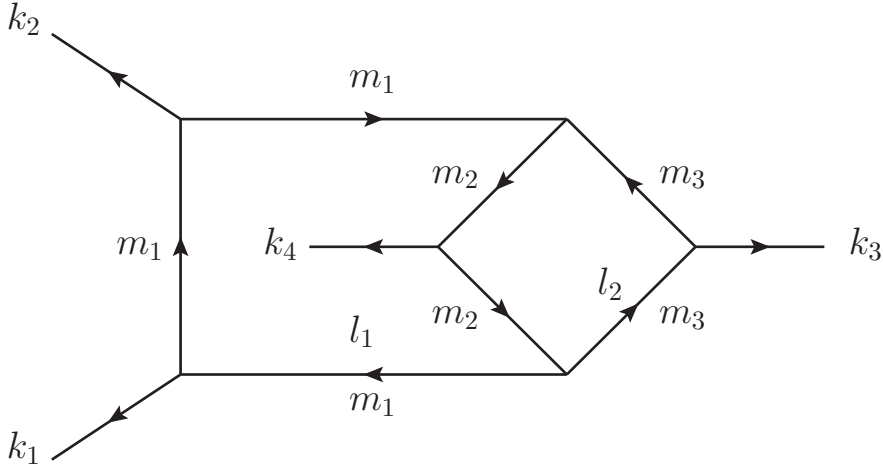}
\end{center}
\caption{Nonplanar double box}\label{crossdia}  
\end{figure}

We now proceed in studying the integral reduction of the massive
nonplanar double box (Fig. \ref{crossdia}). Unlike the previous
examples, this diagram's maximal unitarity cut provides a genus-$3$
hyperelliptic curve \cite{Huang:2013kh, Hauenstein:2014mda}. The structure of holomorphic/meromorphic forms on
this curve is different from the elliptic case. However, our new
approach for integral reduction works for this case as well.

To illustrate our method, we consider the two-loop crossed box
diagrams with massless external legs and three internel masses scales
$\{m_1,m_2,m_3\}$. Our method also works for other crossed box
configurations with all massive internal legs.

\subsection{Maximal Unitarity and geometric properties}
The denominators for the Fig. \ref{crossdia} are,
 \begin{equation}
\label{denom}
\begin{split}
D_1 &=l_1^2-m_1^2 \,,
\qquad
D_2 =(l_1-k_1)^2-m_1^2  \,,
\qquad
D_3 =(l_1-k_1-k_2)^2-m_1^2  \,,
\\
D_4 &=l_2^2-m_3^2 \,,
\qquad
D_5 =(l_2-k_3)^2-m_3^2  \,,
\qquad
D_6 =(l_1-l_2+k_4)^2-m_2^2  \,,
\\
D_7&=(l_1+l_2)^2-m_2^2 \,.
\end{split}
\end{equation}
The
on-shell constrains are 
\begin{equation}
  \label{eq:20}
  D_1=\ldots=D_7=0,
\end{equation}
We use the same loop momenta parametrization (\ref{dbox_Para}). Again, we first solve
for $\alpha_1$, $\alpha_2$, $\alpha_3$, $\beta_1$, $\beta_2$ and
$\beta_4$ in terms of $\alpha_4$ and $\beta_3$,
 \begin{equation}
\label{solutions}
\begin{split}
\alpha_1 &= 1 \,,
\qquad
\alpha_2 = 0 \,,
\qquad
\alpha_3 = \frac{m_1^2  t (s + t)}{
 \alpha_4 s^3},\\
\beta_1 &=-(\alpha_4+\alpha_3+\frac{t}{s})\,,
\qquad
 \beta_2 = 0\,, \qquad
 \beta_4 =\frac{(m_3^2) t (s + t)}{
 \beta_3 s^3}\,.
\end{split}
\end{equation}
The rest two variables satisfy a polynomial equation,
\begin{equation}
  K(\alpha_4,\beta_3)=A(\alpha_4) \beta_3^2+B(\alpha_4)
  \beta_3+C(\alpha_4) =0,
\label{hyperelliptic_xbox}
\end{equation}
whose solution can be formally represented as,
\begin{equation}
  \label{eq:21}
   \beta_3 = \frac{-B(\alpha_4) \pm
     \sqrt{\Delta(\alpha_4)}}{2A(\alpha_4)},\quad \Delta\equiv B^2-4AC 
\end{equation}
Unlike the previous examples, $\Delta(\alpha_4)$ here is a degree-$8$
polynomial in $\alpha_4$ with $8$ distinct roots. Hence the unitarity cut of this diagram
provides a genus-$3$ hyperelliptic curve. (See the appendix for the
classification of complex algebraic curves). Note that not all genus-3
algebraic curves are hyperelliptic, but this one is
because of  (\ref{hyperelliptic_xbox}).

Before the integral reduction, it is interesting to the see the
geometric properties of this unitarity cut.  Using
(\ref{hyperelliptic_xbox}) and the statements of appendix, we see that the function
$\alpha_4$ is a meromorphic function of degree $2$ on the curve, i.e.,
it has two poles $P_1$, $P_2$ and two zeros $Q_1$, $Q_2$. Explicitly,
we can check that $Q_1$ and $Q_2$ are distinct, therefore, in the
language of divisors (\ref{divisor_f}),
\begin{equation}
(\alpha_4)=Q_1+Q_2-P_1-P_2\,.
\end{equation}
The divisor of $\alpha_3$ is then
\begin{equation}
(\alpha_3)=P_1+P_2-Q_1-Q_2\,.
\end{equation} 
The divisor for the function $\beta_3$ and $\beta_4$ is  more
complicated. From  (\ref{hyperelliptic_xbox}), we determined that $\beta_3$
on the cut becomes a meromorphic functions of $4$ simple poles.
The divisor of $\beta_3$ is:
\begin{equation}
(\beta_3)=P_2+Q_2+W_1+W_2-P_1-Q_1-Z_1-Z_2\,.
\end{equation} 
We find that two poles of $\alpha_4$ become a pole and a zero of
$\beta_3$.  Similarly, the two zeros of $\alpha_4$ also become a pole and a zero of
$\beta_3$. 
The divisor of $\beta_4$ is:
\begin{equation}
(\beta_4)=P_1+Q_1+Z_1+Z_2-P_2-Q_2-W_1-W_2\,.
\end{equation}
In summary, there are $8$ poles on this hyperelliptic curve from
numerators insertions, namely
$P_1$, $P_2$, $Q_1$, $Q_2$, $Z_1$, $Z_2$, $W_1$ and $W_1$.


\subsection{Integral reduction}\label{Inum}
First, the integrand reduction via Gr\"obner basis \cite{Zhang:2012ce,Mastrolia:2012an} determines that,
the integrand basis contains $38$ terms in the numerator,
\begin{equation}\label{basisv}
\begin{split}
\{ &\alpha_3^5 \beta_3,\alpha_3^6,\alpha_4^5 \beta_3,\alpha_4^6,\alpha_3^4 \beta_3,\alpha_3^5,\alpha_4 \beta_4^4,\alpha_4
   \beta_3^4,\alpha_4^4 \beta_3,\alpha_4^5,\beta_4^4,\beta_3^4,\alpha_3^3 \beta_3,\alpha_3^4,\alpha_4 \beta_4^3,  \\
   & \alpha_4
   \beta_3^3,\alpha_4^3 \beta_3,\alpha_4^4,\beta_4^3,\beta_3^3,\alpha_3^2 \beta_3,\alpha_3^3,\alpha_4 \beta_4^2,\alpha_4
   \beta_3^2,\alpha_4^2 \beta_3,\alpha_4^3,\beta_4^2,\beta_3^2,\alpha_3 \beta_3,\alpha_3^2, \\
   & \alpha_4 \beta_4,\alpha_4
   \beta_3,\alpha_4^2,\beta_4,\beta_3,\alpha_3,\alpha_4,1 \}
\end{split}
\end{equation}
Then, consider the maximal cut for the scalar integral of this
diagram. The residue computation gives,  
\begin{equation}
I_\text{xbox}[1]|_{7-cut} = \frac{s^3 (s + t)}{16} \oint \frac{\alpha_4 d \alpha_4}{\sqrt{\Delta(\alpha_4)}}\,.
\label{xbox_scalar}
\end{equation}
Note that unlike the elliptic case, on a genus-3 curve there are
three holomorphic $1$-forms from (\ref{hol_basis}), (which have no pole on the hyperelliptic curve),
\begin{equation}
  \label{eq:24}
 \frac{d \alpha_4}{\sqrt{\Delta(\alpha_4)}},\quad   \frac{\alpha_4 d
   \alpha_4}{\sqrt{\Delta(\alpha_4)}}, \quad  \frac{\alpha_4^2 d \alpha_4}{\sqrt{\Delta(\alpha_4)}}
\end{equation}
the scalar integral cut corresponds to the second one, while
\begin{eqnarray}
  \label{eq:25}
  I_\text{xbox}[\alpha_4]|_{7-cut} \propto \oint
  \frac{\alpha_4^2 d \alpha_4}{\sqrt{\Delta(\alpha_4)}}\,,\quad 
I_\text{xbox}[\alpha_3]|_{7-cut}\propto \oint \frac{ d \alpha_4}{\sqrt{\Delta(\alpha_4)}}\,.
\end{eqnarray}
It is curious that for a crossed box integral with the numerator linear in $l_1$, the
maximal cut always gives a holomorphic 1-form. 

This hyperelliptic curve have $6$ fundamental cycles and $8$ poles as shown
in the previous subsection. By global residue theorem, only $7$ poles'
residues are independent. Therefore we may perform maximal unitarity
by computing integrals over $6+7=13$ contours. Since here we only have
one unitarity cut solution, the number of master integers must be
less than or equal $13$. However, in the following discussion, for the
purpose of the integral reduction, we use our new approach to find
exact meromorphic forms instead of calculating these integrals
explicitly.

Following what we did for elliptic cases, we would like to generate the IBP
relations by exact meromorphic 1-forms on the hyperelliptic
curve. Again, we calculate the differential forms $\{d \alpha_3,d \alpha_4,d \beta_3 ,d \beta_4 \}$.
\begin{equation}
d\alpha_4=\frac{\eta}{\eta} d\alpha_4=(2A(\alpha_4) \beta_4+B(\alpha_4)) \frac{d\alpha_4}{\eta}= (2A(\alpha_4) \beta_4-B(\alpha_3)) \frac{\alpha_3}{\lambda_1} \frac{\alpha_4~d\alpha_4}{\eta}\,,
\end{equation}
where we have used the usual definition $\eta\equiv \sqrt{\Delta}$. In the
second equality, we used the on-shell identity,
\begin{equation}
  \label{eq:26}
 \alpha_3\alpha_4 = \lambda_1 \equiv \frac{m_1^2 t (s+t)}{s^3},
\end{equation}
to recover the form of the scalar integral cut (\ref{xbox_scalar}). The step is not needed
for elliptic cases. Then,
\begin{equation}
d\alpha_3=d\left(\frac{\lambda_1}{\alpha_4}\right)=-\lambda_1 \frac{1}{ \alpha_4^2}d\alpha_4=
-\frac{\alpha_3^2}{\lambda_1} d\alpha_4\,,
\end{equation}
where again we have used (\ref{solutions}) to simplify our
expression. The exterior derivatives for $\beta_i$ are more complicated, 
\begin{equation}
d\beta_3=-\left(A^{\prime}(\alpha_4)\beta_3^2+B^{\prime}(\alpha_4)\beta_3+C^{\prime}(\alpha_4)
\right)\frac{\alpha_3}{\lambda_1}\frac{\alpha_4 d\alpha_4}{\eta}\, ,
\end{equation}
and,
\begin{equation}
d\beta_4=d\left(\frac{\lambda_2}{\beta_3} \right)=-\lambda_2
\frac{1}{\beta_3^2}d\beta_3=-\frac{\beta_4^2}{\lambda_2}d\beta_3\, ,
\quad \lambda_2=\frac{m_3^2 t(s+t)}{s^3}
\end{equation}
Given a polynomial function of $\{\alpha_i,\beta_i\}$, we can use the chain rule
to generate the on-shell IBPs.  

We also consider Levi-Civita insertions
which are integrated to zero, 
\begin{equation}\label{omegal2}
\begin{split}
&\epsilon(l_2,k_2,k_3,k_4) ~  l_2 \cdot k_1\,, \qquad \epsilon(l_2,k_2,k_3,k_4)   ~ l_1 \cdot k_4\,, \qquad \epsilon(l_1,l_2,k_1,k_2)\,, \\
 &\epsilon(l_1,l_2,k_1,k_3)\,, \qquad \epsilon(l_1,k_2,k_3,k_4)\,, \qquad \epsilon(l_2,k_2,k_3,k_4)\,.
\end{split}
\end{equation}
In total, we generate $25$ on-shell IBPs and $6$ Levi-Civita
insertions identities. Hence there are $38-25-6=7$ MIs for the non-planar
crossed box diagram with three internal mass scales. Define that
$X=(l_1+ p_4)^2/2$ and $Y=(l_2+ p_1)^2/2$, and 
the MIs can be chosen as:
\begin{equation}
\left\{I_\text{xbox}[X^3],I_\text{xbox}[Y^2],I_\text{xbox}[X Y],I_\text{xbox}[X^2],I_\text{xbox}[X],I_\text{xbox}[Y],I_\text{xbox}[1]    \right\}\,.
\end{equation}  
For this non-planar diagram, the analytic integral reduction relations are
significantly more complicated. For example,
\begin{gather}
I_\text{xbox} \left[ Y^3\right] = \frac{I_\text{xbox} \left[ X^3\right] (28 m_1^2 s + 2 m_3^2 s - 7 s^2 + 4 m_3^2 t - 
   2 m_2^2 (s + 2 t))}{8 (4 m_1^2 - s) s}+\nonumber \\ +\frac{3 I_\text{xbox} \left[ Y^2 \right] (4 m_1^2 (4 m_3^2 - s - 2 t) + s (-2 m_2^2 - 2 m_3^2 + s + 2 t))}{32 m_1^2 - 8 s}+ \nonumber \\
 - \frac{I_\text{xbox} \left[ X Y\right] (4 m_1^2 (s + 2 t) - s (6 m_2^2 - 6 m_3^2 + s + 2 t))}{32 m_1^2 - 8 s}+ \nonumber \\ - \frac{I_\text{xbox} \left[ X^2 \right] (3 m_1^2 - s + t) (28 m_1^2 s + 2 m_3^2 s - 7 s^2 + 4 m_3^2 t - 
    2 m_2^2 (s + 2 t))}{16 (4 m_1^2 - s) s}\nonumber 
\end{gather}
\begin{gather}
-\frac{1}{16 (4 m_1^2 - s) s}
 I_\text{xbox} \left[ Y \right] (4 m_1^4 (3 s^2 + 2 s t + 4 t^2) + 
    s (-2 m_2^4 s - 2 m_3^4 s - s t (3 s + 2 t) + \nonumber \\ + 
       m_3^2 (3 s^2 + 4 s t + 4 t^2) + 
       m_2^2 (-8 m_3^2 s + 3 s^2 + 16 s t + 4 t^2)) + 
    m_1^2 (48 m_3^4 s +\nonumber \\ + s (-3 s^2 + 10 s t + 4 t^2) - 
       2 m_2^2 (s^2 + 20 s t + 8 t^2) - 
       2 m_3^2 (7 s^2 + 20 s t + 8 t^2)))+ \nonumber \\ +\frac{1}{32 (4 m_1^2 - s) s}I_\text{xbox} \left[ X \right] (84 m_1^6 s + 
   m_1^4 (-49 s^2 + 24 s t - 32 t^2 - 6 m_2^2 (s + 2 t) + \nonumber \\ + 
      6 m_3^2 (s + 2 t)) + 
   s (-2 m_2^4 s + s (10 m_3^4 + 5 m_3^2 s + 7 s t) + 
      m_2^2 (-8 m_3^2 s + 5 s^2 + 4 s t + 8 t^2))+ \nonumber \\ - 
   m_1^2 (s (-7 s^2 + 34 s t - 8 t^2) + 
      2 m_3^2 (11 s^2 + 8 s t + 4 t^2) + 2 m_2^2 (5 s^2 + 12 t^2)))+
      \nonumber 
\end{gather}
\begin{gather}
- \frac{1}{64 (4 m_1^2 - s) s}I_\text{xbox}[1] (28 m_1^8 s -m_1^6 (7
s^2 + 4 s t + 32 t^2 + 2 m_2^2 (s + 2 t) - 2 m_3^2 (s + 2 t))+
\nonumber \\ -m_1^4 (2 m_2^2 (6 s^2 + s t + 10 t^2) - t (25 s^2 + 48 s
t + 16 t^2) + m_3^2 (52 s^2 + 46 s t + 44 t^2))+ \nonumber \\ - m_1^2
(32 m_3^6 s + 2 m_2^4 s^2 + 2 s t (3 s^2 + 5 s t + 2 t^2) - 2 m_3^4
(s^2 + 16 s t + 16 t^2) +  m_2^2 (-5 s^3 + \nonumber \\ + 38 s^2 t +
24 s t^2 + 16 t^3) + m_3^2 (-13 s^3 + 14 s^2 t + 8 s t^2 + 16 t^3 - 8
m_2^2 (s^2 + 10 s t + 4 t^2)))+ \nonumber \\ + s (m_3^2 t (10 m_3^2 s
+ 5 s^2 + 2 s t + 4 t^2) + 2 m_2^4 (2 m_3^2 s - t (5 s + 4 t)) + m_2^2
(4 m_3^4 s + t (13 s^2 + \nonumber \\ + 10 s t + 4 t^2) - 2 m_3^2 (3
s^2 + 14 s t + 4 t^2))))+\ldots 
\label{xbox_eample}
\end{gather}
where $\ldots$ stands for integrals with few than $7$
propagators. The integral reduction at the level of crossed box
takes about $22$ minutes with our Mathematica code.  We also performed the
integral reduction of crossed box with doubled propagators, by the
same method for the double box case as (\ref{dbox_double_reduction}). All integral reduction
relations obtained by our method have been
numerically verified by \rm{FIRE} \cite{Smirnov:2013dia,Smirnov:2008iw}.

\section{Conclusions}
In this paper, we present the relation between the on-shell IBPs and
the meromorphic one-forms on algebraic curves, for a class of two-loop
diagrams: $D$-dimensional $L$-loop diagram with $DL-1$ propagators and
one unitarity cut solution.  
In this case, the unitarity cut has a globally
well-defined one-dimensional complex structure on it, and hence the analysis of IBPs
relations can be translated into the analysis of complex curves. 

By presenting several two-loop examples, planar and non-planar, we show that from the knowledge of
algebraic curves, it is very easy to construct an IBP relation which
reduces an arbitrary integral to master integrals. No explicit form of
elliptic/hyperelliptic function is needed in our method, since only
the differential relations of these functions are needed. Our method
works for the reduction of integrals with or without doubled propagators.

There are several interesting future directions. In this paper, we
mainly consider diagrams of $DL-1$ propagators without three-point massless vertice. If a
$(DL-1)$-propagator diagram has three-point massless vertices,
generically, the unitarity cut is not an irreducible curve but a
reducible curve, i.e, union of several irreducible algebraic curves. We find that
the IBPs obtained from our method, has a smooth massless limit and the
limit forms a subset of IBPs for these diagrams. For example, the
massless limit of our method, applied on the massless double box diagram,
provides all but $2$ IBPs. The missing $2$
IBPs contain only low-rank numerators, so can be easily found by other
algorithms. So in these cases, our method would greatly speed up the
integral reduction process, even if IBPs are not all obtained. In the
future, we expect that the algebraic geometry analysis on reducible
curves will lead the complete set of on-shell IBPs of these diagrams.

Furthermore, we may consider using geometric properties of {\it algebraic
surfaces} to study loop diagrams with an arbitrary number of
propagators. It is well known that the algebraic geometry property
of surfaces
is more complicated than that for curves. However, we expect our
approach will be generalized for the surfaces cases, because essentially
our method does not depend on the detailed information of
elliptic/hyperelliptic functions or integrals. Only the complex
structure and differential relations are needed. So the surface cases
would be studied following this direction, and to recover the ``$\ldots$''
terms in our reduction like (\ref{dbox_eample}) and (\ref{xbox_eample}).

Finally, we will study the $\epsilon$-dependent part of the integral
reduction, based on our method. In this paper, we consider diagrams
with 
integer-valued spacetime dimension. The ongoing research on two-loop maximal
unitarity in dimensional regularization scheme \cite{2loopDMU}, also based on
algebraic geometry tools, will help us to understand IBPs with dimensional
regularization from a geometric viewpoint. 

\acknowledgments
We would like to express our sincere gratitude to Niklas Beisert, for
his advices and help from the beginning of this project. We also thank Simon
Badger, Johannes Br\"odel, Lance Dixon, Hjalte Frellesvig, Johannes
Henn, David Kosower, Kasper Larsen, Lorenzo Tancredi, Stefan Weinzierl,
Congkao Wen for enlightening discussions on this research
direction. Especially, we thank Stefan	M\"uller-Stach for his
valuable intuitions on complex geometry related to integral reduction, Mads S\o gaard for his participation during the early stage of
this project, and Benjamin Page for his careful reading of our draft
and suggestions. YZ is grateful for Mainz Institute for Theoretical
Physics for the hospitality in the scientific program ``Amplitudes, Motives
and Beyond'' and its partial support during the
completion of the work.

The research leading to results in this paper, has received
funding from the European Research Council under the European Union's
Seventh Framework Programme (FP/2007-2013) / ERC Grant Agreement
no. 615203. The work is partially supported by the Swiss National Science Foundation through the NCCR SwissMAP.

\appendix
\section{Rudiments of Algebraic Curves}\label{riemann}
In this appendix we give a brief introduction to the mathematical
background of algebraic curves used in this paper. The extensive
treatment can be found in ref. \cite{farkas1992riemann,
  miranda1995algebraic, griffiths2011principles}.

\label{riesurf}

\begin{mydef}\label{d1}
A Riemann surface is a connected one-dimensional complex manifold.
\end{mydef}

We are mostly interested in compact Riemann surface. Any compact Riemann surface is homeomorphic to a
sphere with $g\geq 0$ handles attached, and the number $g$ is called the genus of the
Riemann surface. Since the complex dimension is one, we also
denote a compact Riemann surface as a complex algebraic
curve. However, rigorously speaking, we may need to blow up possible
singular points on an algebraic curve to make it a Riemann
surface.

\begin{mydef}\label{d2}
A holomorphic map between Riemann surfaces $X$ and $Y$ is a continuous map $f : X \rightarrow Y$ such that for each holomorphic coordinate $\phi_U$ on $U$ containing $x$ on X and $\psi_W$ defined in a neighbourhood of $f(x)$ on $Y$, the composition
\begin{equation}
\psi_W \circ f \circ \phi_U^{-1}
\end{equation}
is holomorphic.
\end{mydef}

\begin{mydef}
A meromorphic function $f$ on a Riemann surface X is a holomorphic map to the Riemann sphere $S = \mathbb{C} \cup \{\infty\}$.
\end{mydef}

One very useful theorem regarding to the topological properties of
algebraic curves is Riemann-Hurwitz theorem. Here we consider the
special cases of $f: X \rightarrow S$. If in a neighborhood of $P\in
X$, $P$ is located at the origin and $f$ has the expansion $f(z) -f(0)\sim
z^n$, $n>1$, then we say $P$ is a {\it ramified point} of $f$ and $n$
is the ramification index of $P$.  Removing images of {\it ramification points} under $f$,
we get a Riemann sphere excluding a finite number of points, namely
$\hat S$. For any
point $Q\in \hat S$, define that $d(Q)\equiv \#f^{-1}(Q)$, the number
of points in the inverse image. $d$ is an integer-valued and
continuous function, hence a constant. This constant $d$ is called the degree of $f$.

\begin{mytheo}\label{RH1}(Riemann-Hurwitz)
Let $f : X\rightarrow S$ be a meromorphic function of degree $d$ on a
closed connected Riemann surface $X$. The ramified points are
$x_1,\ldots, x_n$, with multiplicity $m_1,\ldots, m_n$ . Then the
Euler character of $X$,
\begin{equation}\label{rhf}
\chi (X)=2d-\sum_{k=1}^n (m_k-1).
\end{equation}
\end{mytheo}
For a compact Riemann surface, the Euler character is related to the
genus $g$, i.e., number of handles as,
\begin{equation}
  \label{eq:27}
  \chi=2-2g .
\end{equation}

The $g=0$ compact Riemann surface is a Riemann sphere, while $g=1$
compact Riemann surface is an elliptic curve (or torus
topologically). $g>1$ cases are more complicated, and we focus a
particular class, {\it hyperelliptic curves}, which is defined as an algebraic curve,
\begin{equation}
  \label{hyperelliptic}
 \mathcal C: y^2=h(x)
\end{equation}
$h$ is a degree-$n$ polynomial in $x$ with $n$ distinct
roots. $\mathcal C$ has the genus $g$, if $d=2g+1$ or $d=2g+2$, by (\ref{rhf}). Note
that $g=2$ curve must be hyperelliptic, but not all $g>2$ curves are hyperelliptic.

In the hyperelliptic case, $x$ and $y$ become
meromorphic function with these properties, 
\begin{itemize}\label{hypsb}
\item $x$ is a meromorphic function of degree $2$ on $\mathcal C$,
\item $y$ is a meromorphic function of degree $n$ on $\mathcal C$,
\item $x:$ $\mathcal C \rightarrow S$ has $2g+2$ ramified points.  If
  $n$ is even, these points are $(x,y)=(a_i,0)$ where $a_i$'s are the roots
  of $h(x)$. If $n$ is odd, these points are $(x,y)=(a_i,0)$ and the
  point at infinity.
\item  Every meromorphic function $f$ on $\mathcal C$ can be uniquely written as
\begin{equation}
f=r(x)+y s(x)
\label{generators_hyperelliptic}
\end{equation}
where $r(x)$ and $s(x)$ are rational functions of x.
\end{itemize}
The last property (\ref{generators_hyperelliptic}) is important for studying the exact meromorphic
1-forms, which play the central role of our integral reduction
algorithm. 

\subsection{Riemann-Roch theorem}\label{r-r}
We now want to state one of the fundamental theorems of compact
Riemann surface $X$. First, we present several definitions,
\begin{mydef}
A divisor D on a compact Riemann surface X is a formal sum of points with multiplicities.
\begin{equation}
D=\sum_{P\in X} n_p P,
\end{equation}
$D\geq 0$ if and only if $n_P \geq 0$, $\forall P \in X$.
\end{mydef}
We can naturally associate a divisor to a meromorphic function $f$ in the following way,
\begin{equation}
(f)=\sum_{P \in X}(\text{ord}_P(f)) P.
\label{divisor_f}
\end{equation}
where $\text{ord}_P(f)$ is the leading power of $f$'s Laurent expansion at
$P$. $\deg(D)$ is the degree of the divisor defined as
\begin{equation}
\deg(D)=\sum_{P \in X} n_P\,.
\end{equation}
$\mathcal L(D)$ is the space of meromorphic functions $f$ for which,
$(f)+D\geq 0$. We define $l(D)=\dim \mathcal L(D)$. Let $K$ be the canonical divisor associated with any meromorphic one form, 
\begin{equation}
  \label{eq:18}
  i(D)\equiv l(K-D).
\end{equation}
We can now state the theorem:
\begin{mytheo}\label{rr}
(Riemann-Roch) Let $X$ be a compact Riemann surface of genus $g$ and
$D\in$ a divisor.  
\begin{equation}\label{rre}
l(D)-i(D)=\deg(D)-g+1\,.
\end{equation}
\end{mytheo}
It is clear that If $D < 0$ then $l(D) = 0$, so the Riemann-Roch
theorem simplifies as,
\begin{equation}\label{r-r1}
i(D)=-\deg(D)+g-1\,.
\end{equation}
On the other hand, if $\deg D \geq 2g-2$ then $i(D) = 0$. We have
\begin{equation}
l(D)=\deg(D)-g+1
\end{equation}

\begin{mytheo}
If X is a compact Riemann surface of genus $g$ then
\begin{enumerate}
\item The space of holomorphic one-forms on $X$ form a finite dimensional vector
space of complex dimension $g$,
\item If $\omega$ is a meromorphic differential on a Riemann surface X then the number
of zeros of $\omega$ minus the number of poles, counted with multiplicity is $2g- 2$.
\end{enumerate}
\end{mytheo}

For the hyperelliptic curve (\ref{hyperelliptic}), we can find an explicit basis of the holomorphic one-forms,
\begin{mycorol}
The $g$ differentials 
\begin{equation}
\label{hol_basis}
\frac{x^j dx}{y}\,, \qquad j=0,\ldots,g-1\,,
\end{equation}
form a basis of holomorphic differential forms.
\end{mycorol}
We use this basis frequently in our paper for the integral reduction.


\bibliographystyle{unsrt}

\bibliography{2loop_AC}

\begin{thebibliography}{10}

\bibitem{Chetyrkin:1981qh}
K.G. Chetyrkin and F.V. Tkachov.
\newblock {Integration by Parts: The Algorithm to Calculate beta Functions in 4
  Loops}.
\newblock {\em Nucl.Phys.}, B192:159--204, 1981.

\bibitem{Anastasiou:2004vj}
Charalampos Anastasiou and Achilleas Lazopoulos.
\newblock {Automatic integral reduction for higher order perturbative
  calculations}.
\newblock {\em JHEP}, 0407:046, 2004.

\bibitem{Smirnov:2013dia}
A.V. Smirnov and V.A. Smirnov.
\newblock {FIRE4, LiteRed and accompanying tools to solve integration by parts
  relations}.
\newblock {\em Comput.Phys.Commun.}, 184:2820--2827, 2013.

\bibitem{Smirnov:2008iw}
A.V. Smirnov.
\newblock {Algorithm FIRE -- Feynman Integral REduction}.
\newblock {\em JHEP}, 0810:107, 2008.

\bibitem{Smirnov:2006tz}
A.V. Smirnov.
\newblock {An Algorithm to construct Grobner bases for solving integration by
  parts relations}.
\newblock {\em JHEP}, 0604:026, 2006.

\bibitem{vonManteuffel:2012yz}
A.~von Manteuffel and C.~Studerus.
\newblock {Reduze 2 - Distributed Feynman Integral Reduction}.
\newblock 2012.

\bibitem{Studerus:2009ye}
C.~Studerus.
\newblock {Reduze-Feynman Integral Reduction in C++}.
\newblock {\em Comput.Phys.Commun.}, 181:1293--1300, 2010.

\bibitem{Laporta:2001dd}
S.~Laporta.
\newblock {High precision calculation of multiloop Feynman integrals by
  difference equations}.
\newblock {\em Int.J.Mod.Phys.}, A15:5087--5159, 2000.

\bibitem{Lee:2008tj}
R.N. Lee.
\newblock {Group structure of the integration-by-part identities and its
  application to the reduction of multiloop integrals}.
\newblock {\em JHEP}, 0807:031, 2008.

\bibitem{Gluza:2010ws}
Janusz Gluza, Krzysztof Kajda, and David~A. Kosower.
\newblock {Towards a Basis for Planar Two-Loop Integrals}.
\newblock {\em Phys.Rev.}, D83:045012, 2011.

\bibitem{Schabinger:2011dz}
Robert~M. Schabinger.
\newblock {A New Algorithm For The Generation Of Unitarity-Compatible
  Integration By Parts Relations}.
\newblock {\em JHEP}, 1201:077, 2012.

\bibitem{vonManteuffel:2014ixa}
Andreas von Manteuffel and Robert~M. Schabinger.
\newblock {A novel approach to integration by parts reduction}.
\newblock {\em Phys. Lett.}, B744:101--104, 2015.

\bibitem{Zhang:2014xwa}
Yang Zhang.
\newblock {Integration-by-parts identities from the viewpoint of differential
  geometry}.
\newblock 2014.

\bibitem{Lee:2013hzt}
Roman~N. Lee and Andrei~A. Pomeransky.
\newblock {Critical points and number of master integrals}.
\newblock {\em JHEP}, 11:165, 2013.

\bibitem{Bern:1994zx}
Zvi Bern, Lance~J. Dixon, David~C. Dunbar, and David~A. Kosower.
\newblock {One loop n point gauge theory amplitudes, unitarity and collinear
  limits}.
\newblock {\em Nucl.Phys.}, B425:217--260, 1994.

\bibitem{Bern:1994cg}
Zvi Bern, Lance~J. Dixon, David~C. Dunbar, and David~A. Kosower.
\newblock {Fusing gauge theory tree amplitudes into loop amplitudes}.
\newblock {\em Nucl.Phys.}, B435:59--101, 1995.

\bibitem{Britto:2004nc}
Ruth Britto, Freddy Cachazo, and Bo~Feng.
\newblock {Generalized unitarity and one-loop amplitudes in N=4
  super-Yang-Mills}.
\newblock {\em Nucl.Phys.}, B725:275--305, 2005.

\bibitem{Britto:2004ap}
Ruth Britto, Freddy Cachazo, and Bo~Feng.
\newblock {New recursion relations for tree amplitudes of gluons}.
\newblock {\em Nucl.Phys.}, B715:499--522, 2005.

\bibitem{Britto:2004nj}
Ruth Britto, Freddy Cachazo, and Bo~Feng.
\newblock {Computing one-loop amplitudes from the holomorphic anomaly of
  unitarity cuts}.
\newblock {\em Phys. Rev.}, D71:025012, 2005.

\bibitem{Britto:2005fq}
Ruth Britto, Freddy Cachazo, Bo~Feng, and Edward Witten.
\newblock {Direct proof of tree-level recursion relation in Yang-Mills theory}.
\newblock {\em Phys.Rev.Lett.}, 94:181602, 2005.

\bibitem{Britto:2005ha}
Ruth Britto, Evgeny Buchbinder, Freddy Cachazo, and Bo~Feng.
\newblock {One-loop amplitudes of gluons in SQCD}.
\newblock {\em Phys.Rev.}, D72:065012, 2005.

\bibitem{Anastasiou:2006jv}
Charalampos Anastasiou, Ruth Britto, Bo~Feng, Zoltan Kunszt, and Pierpaolo
  Mastrolia.
\newblock {D-dimensional unitarity cut method}.
\newblock {\em Phys.Lett.}, B645:213--216, 2007.

\bibitem{Anastasiou:2006gt}
Charalampos Anastasiou, Ruth Britto, Bo~Feng, Zoltan Kunszt, and Pierpaolo
  Mastrolia.
\newblock {Unitarity cuts and Reduction to master integrals in d dimensions for
  one-loop amplitudes}.
\newblock {\em JHEP}, 0703:111, 2007.

\bibitem{Cachazo:2004dr}
Freddy Cachazo.
\newblock {Holomorphic anomaly of unitarity cuts and one-loop gauge theory
  amplitudes}.
\newblock 2004.

\bibitem{Kosower:2011ty}
David~A. Kosower and Kasper~J. Larsen.
\newblock {Maximal Unitarity at Two Loops}.
\newblock 2011.

\bibitem{Johansson:2012zv}
Henrik Johansson, David~A. Kosower, and Kasper~J. Larsen.
\newblock {Two-Loop Maximal Unitarity with External Masses}.
\newblock {\em Phys.Rev.}, D87:025030, 2013.

\bibitem{Sogaard:2013yga}
Mads Sogaard.
\newblock {Global Residues and Two-Loop Hepta-Cuts}.
\newblock 2013.

\bibitem{Johansson:2013sda}
Henrik Johansson, David~A. Kosower, and Kasper~J. Larsen.
\newblock {Maximal Unitarity for the Four-Mass Double Box}.
\newblock 2013.

\bibitem{CaronHuot:2012ab}
Simon Caron-Huot and Kasper~J. Larsen.
\newblock {Uniqueness of two-loop master contours}.
\newblock {\em JHEP}, 1210:026, 2012.

\bibitem{Sogaard:2013fpa}
Mads S{\o}gaard and Yang Zhang.
\newblock {Multivariate Residues and Maximal Unitarity}.
\newblock {\em JHEP}, 12:008, 2013.

\bibitem{Sogaard:2014jla}
Mads S{{\o{}}}gaard and Yang Zhang.
\newblock {Elliptic Functions and Maximal Unitarity}.
\newblock {\em Phys.Rev.}, D91(8):081701, 2015.

\bibitem{Johansson:2015ava}
Henrik Johansson, David~A. Kosower, Kasper~J. Larsen, and Mads S{\o}gaard.
\newblock {Cross-Order Integral Relations from Maximal Cuts}.
\newblock {\em Phys. Rev.}, D92(2):025015, 2015.

\bibitem{griffiths2011principles}
P.~Griffiths and J.~Harris.
\newblock {\em Principles of Algebraic Geometry}.
\newblock Wiley Classics Library. Wiley, 2011.

\bibitem{miranda1995algebraic}
R.~Miranda.
\newblock {\em Algebraic Curves and Riemann Surfaces}.
\newblock Dimacs Series in Discrete Mathematics and Theoretical Comput.
  American Mathematical Society, 1995.

\bibitem{farkas1992riemann}
H.M. Farkas and I.~Kra.
\newblock {\em Riemann Surfaces: With 27 Figures}.
\newblock Graduate Texts in Mathematics. Springer New York, 1992.

\bibitem{Ossola:2006us}
Giovanni Ossola, Costas~G. Papadopoulos, and Roberto Pittau.
\newblock {Reducing full one-loop amplitudes to scalar integrals at the
  integrand level}.
\newblock {\em Nucl.Phys.}, B763:147--169, 2007.

\bibitem{Ossola:2007ax}
Giovanni Ossola, Costas~G. Papadopoulos, and Roberto Pittau.
\newblock {CutTools: A Program implementing the OPP reduction method to compute
  one-loop amplitudes}.
\newblock {\em JHEP}, 0803:042, 2008.

\bibitem{Mastrolia:2011pr}
Pierpaolo Mastrolia and Giovanni Ossola.
\newblock {On the Integrand-Reduction Method for Two-Loop Scattering
  Amplitudes}.
\newblock {\em JHEP}, 1111:014, 2011.

\bibitem{Badger:2012dv}
Simon Badger, Hjalte Frellesvig, and Yang Zhang.
\newblock {An Integrand Reconstruction Method for Three-Loop Amplitudes}.
\newblock {\em JHEP}, 1208:065, 2012.

\bibitem{Zhang:2012ce}
Yang Zhang.
\newblock {Integrand-Level Reduction of Loop Amplitudes by Computational
  Algebraic Geometry Methods}.
\newblock {\em JHEP}, 1209:042, 2012.

\bibitem{Mastrolia:2012an}
Pierpaolo Mastrolia, Edoardo Mirabella, Giovanni Ossola, and Tiziano Peraro.
\newblock {Scattering Amplitudes from Multivariate Polynomial Division}.
\newblock {\em Phys.Lett.}, B718:173--177, 2012.

\bibitem{Feng:2012bm}
Bo~Feng and Rijun Huang.
\newblock {The classification of two-loop integrand basis in pure
  four-dimension}.
\newblock {\em JHEP}, 02:117, 2013.

\bibitem{Mastrolia:2012wf}
Pierpaolo Mastrolia, Edoardo Mirabella, Giovanni Ossola, and Tiziano Peraro.
\newblock {Integrand-Reduction for Two-Loop Scattering Amplitudes through
  Multivariate Polynomial Division}.
\newblock {\em Phys.Rev.}, D87:085026, 2013.

\bibitem{Mastrolia:2013kca}
Pierpaolo Mastrolia, Edoardo Mirabella, Giovanni Ossola, and Tiziano Peraro.
\newblock {Multiloop Integrand Reduction for Dimensionally Regulated
  Amplitudes}.
\newblock {\em Phys.Lett.}, B727:532--535, 2013.

\bibitem{Badger:2013gxa}
Simon Badger, Hjalte Frellesvig, and Yang Zhang.
\newblock {A Two-Loop Five-Gluon Helicity Amplitude in QCD}.
\newblock {\em JHEP}, 12:045, 2013.

\bibitem{Huang:2013kh}
Rijun Huang and Yang Zhang.
\newblock {On Genera of Curves from High-loop Generalized Unitarity Cuts}.
\newblock {\em JHEP}, 1304:080, 2013.

\bibitem{Hauenstein:2014mda}
Jonathan~D. Hauenstein, Rijun Huang, Dhagash Mehta, and Yang Zhang.
\newblock {Global Structure of Curves from Generalized Unitarity Cut of
  Three-loop Diagrams}.
\newblock {\em JHEP}, 02:136, 2015.

\bibitem{Sogaard:2014ila}
Mads Sogaard and Yang Zhang.
\newblock {Unitarity Cuts of Integrals with Doubled Propagators}.
\newblock {\em JHEP}, 1407:112, 2014.

\bibitem{Sogaard:2014oka}
Mads Sogaard and Yang Zhang.
\newblock {Massive Nonplanar Two-Loop Maximal Unitarity}.
\newblock {\em JHEP}, 12:006, 2014.

\bibitem{Broadhurst:1993mw}
David~J. Broadhurst, J.~Fleischer, and O.~V. Tarasov.
\newblock {Two loop two point functions with masses: Asymptotic expansions and
  Taylor series, in any dimension}.
\newblock {\em Z. Phys.}, C60:287--302, 1993.

\bibitem{Berends:1993ee}
Frits~A. Berends, M.~Buza, M.~Bohm, and R.~Scharf.
\newblock {Closed expressions for specific massive multiloop selfenergy
  integrals}.
\newblock {\em Z. Phys.}, C63:227--234, 1994.

\bibitem{Bauberger:1994nk}
S.~Bauberger, M.~Bohm, G.~Weiglein, Frits~A. Berends, and M.~Buza.
\newblock {Calculation of two loop selfenergies in the electroweak standard
  model}.
\newblock {\em Nucl. Phys. Proc. Suppl.}, 37B:95--114, 1994.

\bibitem{Bauberger:1994by}
S.~Bauberger, Frits~A. Berends, M.~Bohm, and M.~Buza.
\newblock {Analytical and numerical methods for massive two loop selfenergy
  diagrams}.
\newblock {\em Nucl. Phys.}, B434:383--407, 1995.

\bibitem{Bauberger:1994hx}
S.~Bauberger and M.~Bohm.
\newblock {Simple one-dimensional integral representations for two loop
  selfenergies: The Master diagram}.
\newblock {\em Nucl. Phys.}, B445:25--48, 1995.

\bibitem{Caffo:1998du}
Michele Caffo, H.~Czyz, S.~Laporta, and E.~Remiddi.
\newblock {The Master differential equations for the two loop sunrise selfmass
  amplitudes}.
\newblock {\em Nuovo Cim.}, A111:365--389, 1998.

\bibitem{Groote:1998wy}
S.~Groote, J.~G. Korner, and A.~A. Pivovarov.
\newblock {On the evaluation of sunset - type Feynman diagrams}.
\newblock {\em Nucl. Phys.}, B542:515--547, 1999.

\bibitem{Groote:1998ic}
S.~Groote, J.~G. Korner, and A.~A. Pivovarov.
\newblock {A New technique for computing the spectral density of sunset type
  diagrams: Integral transformation in configuration space}.
\newblock {\em Phys. Lett.}, B443:269--275, 1998.

\bibitem{Caffo:2002ch}
Michele Caffo, H.~Czyz, and E.~Remiddi.
\newblock {Numerical evaluation of the general massive 2 loop sunrise selfmass
  master integrals from differential equations}.
\newblock {\em Nucl. Phys.}, B634:309--325, 2002.

\bibitem{Laporta:2004rb}
S.~Laporta and E.~Remiddi.
\newblock {Analytic treatment of the two loop equal mass sunrise graph}.
\newblock {\em Nucl. Phys.}, B704:349--386, 2005.

\bibitem{Pozzorini:2005ff}
S.~Pozzorini and E.~Remiddi.
\newblock {Precise numerical evaluation of the two loop sunrise graph master
  integrals in the equal mass case}.
\newblock {\em Comput. Phys. Commun.}, 175:381--387, 2006.

\bibitem{Groote:2005ay}
S.~Groote, J.~G. Korner, and A.~A. Pivovarov.
\newblock {On the evaluation of a certain class of Feynman diagrams in x-space:
  Sunrise-type topologies at any loop order}.
\newblock {\em Annals Phys.}, 322:2374--2445, 2007.

\bibitem{Kniehl:2005bc}
B.~A. Kniehl, A.~V. Kotikov, A.~Onishchenko, and O.~Veretin.
\newblock {Two-loop sunset diagrams with three massive lines}.
\newblock {\em Nucl. Phys.}, B738:306--316, 2006.

\bibitem{Groote:2012pa}
S.~Groote, J.~G. Korner, and A.~A. Pivovarov.
\newblock {A Numerical Test of Differential Equations for One- and Two-Loop
  sunrise Diagrams using Configuration Space Techniques}.
\newblock {\em Eur. Phys. J.}, C72:2085, 2012.

\bibitem{Bailey:2008ib}
David~H. Bailey, Jonathan~M. Borwein, David Broadhurst, and M.~L. Glasser.
\newblock {Elliptic integral evaluations of Bessel moments}.
\newblock {\em J. Phys.}, A41:205203, 2008.

\bibitem{Caffo:2008aw}
Michele Caffo, Henryk Czyz, Michal Gunia, and Ettore Remiddi.
\newblock {BOKASUN: A Fast and precise numerical program to calculate the
  Master Integrals of the two-loop sunrise diagrams}.
\newblock {\em Comput. Phys. Commun.}, 180:427--430, 2009.

\bibitem{Kalmykov:2008ge}
Mikhail~{\relax Yu}. Kalmykov and Bernd~A. Kniehl.
\newblock {Towards all-order Laurent expansion of generalized hypergeometric
  functions around rational values of parameters}.
\newblock {\em Nucl. Phys.}, B809:365--405, 2009.

\bibitem{MullerStach:2011ru}
Stefan M{\"u}ller-Stach, Stefan Weinzierl, and Raphael Zayadeh.
\newblock {A Second-Order Differential Equation for the Two-Loop Sunrise Graph
  with Arbitrary Masses}.
\newblock {\em Commun. Num. Theor. Phys.}, 6:203--222, 2012.

\bibitem{Adams:2013nia}
Luise Adams, Christian Bogner, and Stefan Weinzierl.
\newblock {The two-loop sunrise graph with arbitrary masses}.
\newblock {\em J.Math.Phys.}, 54:052303, 2013.

\bibitem{Remiddi:2013joa}
Ettore Remiddi and Lorenzo Tancredi.
\newblock {Schouten identities for Feynman graph amplitudes; The Master
  Integrals for the two-loop massive sunrise graph}.
\newblock {\em Nucl. Phys.}, B880:343--377, 2014.

\bibitem{Bloch:2013tra}
Spencer Bloch and Pierre Vanhove.
\newblock {The elliptic dilogarithm for the sunset graph}.
\newblock 2013.

\bibitem{Adams:2014vja}
Luise Adams, Christian Bogner, and Stefan Weinzierl.
\newblock {The two-loop sunrise graph in two space-time dimensions with
  arbitrary masses in terms of elliptic dilogarithms}.
\newblock {\em J. Math. Phys.}, 55(10):102301, 2014.

\bibitem{Adams:2015gva}
Luise Adams, Christian Bogner, and Stefan Weinzierl.
\newblock {The two-loop sunrise integral around four space-time dimensions and
  generalisations of the Clausen and Glaisher functions towards the elliptic
  case}.
\newblock 2015.

\bibitem{vanNeerven:1983vr}
W.~L. van Neerven and J.~A.~M. Vermaseren.
\newblock {LARGE LOOP INTEGRALS}.
\newblock {\em Phys. Lett.}, B137:241, 1984.

\bibitem{2loopDMU}
Larsen Kasper and Yang Zhang.
\newblock {Two-loop Maximal Unitarity in Dimensional Regularization}.
\newblock {\em to appear}, 2015.

\end{thebibliography}

\end{document}